\documentstyle[prb,aps,epsf]{revtex}
\begin{document}
\centerline{{\huge A Simple Shell Model for Quantum Dots }}
\centerline{{\huge in a Tilted Magnetic Field} }
\medskip
\centerline{
W.D. Heiss$^{\star}$ and R.G. Nazmitdinov$^{\star,\star \star}$}
\medskip
\begin{center}{
$^{\star}$ Centre for Nonlinear Studies and Department
of Physics
\\University of the Witwatersrand, PO Wits 2050,
Johannesburg, \\South Africa
\\$^{\star \star}$ Bogoliubov Laboratory of Theoretical Physics
\\Joint Institute for Nuclear Research, 141980 Dubna, Russia}
\end{center}
\baselineskip 20pt minus.1pt
\begin{abstract}
A model for quantum dots is proposed, in which
the motion of a few electrons in a three-dimensional harmonic oscillator
potential under the influence of a homogeneous magnetic field of arbitrary
direction is studied. The spectrum and the wave functions are obtained by
solving the classical problem. The ground state of the Fermi-system is obtained
by minimizing the total energy with regard to the confining frequencies.
From this a dependence of the equilibrium shape of the quantum
dot on the electron number,
the magnetic field parameters and the slab thickness is found.
\end{abstract}
\vspace{0.2in}

\vspace{0.2in}

\section{Introduction}
Recent progress in semiconductor technology now allows the fabrication
of artificially structured atoms in semiconductors
called quantum dots, in which
electrons are trapped in a small localized region of space of
a few hundred Angstr\"oms. In these nanostructures the
electron wavelength is of the same length scale as the confinement so that
quantum effects are important, which results in the quantization of
single-electron energy levels with excitation energy of several meV
(see for review \cite{TC,MK,Jon}).
A prominent feature of a Fermi-system which is confined to a finite region of
space is the emergence of shell effects in the single-particle spectrum
as is well known in nuclear physics \cite{BM} and more recently for large
metallic clusters (\cite{Ni}-\cite{Com} and references therein).
In the present
work we address this problem to mesoscopic systems like quantum dots which
contain a small number of electrons.

The electron states of few-electron quantum dots subjected to a strong
magnetic field have been studied extensively in \cite{1}-\cite{6}.
The electrodynamic response of an interacting electron system in the
presence of a confining potential is expected to be dominated by the 
many-body effects of the electrons.
Sikorski and Merkt \cite{1} found experimentally
the surprising result that the resonance frequencies in the magneto-optical
spectrum are independent of the number of electrons in the quantum dot.
The systems which are experimentally realized extent
much less in the the $z-$direction than in the $x-y-$plane.
This has led to a simple description of the 
far-infrared resonance (FIR) frequencies \cite{BJH,P,MC} as the
energy levels of a two-dimensional harmonic oscillator potential
in the presence of a magnetic field \cite{Fock}.
It was interpreted as a consequence of
Kohn's theorem \cite{Kohn} which has been generalized for a parabolic
potential \cite{BJH,P,MC,Bak,Yip,Li}. According to this theorem, the total
Hamiltonian can be divided into two parts, the center-of-mass motion (CM)
and the relative motion which contains the electron-electron interaction.
Since the radiation of an external electric dipole field
couples only to the CM motion and does not affect the relative motion,
the dipole resonance frequencies for the interacting system
should be exactly the same as those of the noninteracting system and be
independent of the electron-electron interaction. 

A full understanding of the
experimental results needs an analysis of many-body effects.
Microscopic calculations using Hartree approximation
for electron numbers $N<10$ \cite{KLS} neglected the exchange
and correlation effects. The direct numerical diagonalization \cite{MC}
have been performed only for three- and four electron systems.
The more complicated resonance structure observed in \cite{3,6}
raised the question as to the validity of Kohn's theorem for 
quantum dots. In order to describe the experimental data
it was assumed that, in real samples, there is a deviation of the confining 
potential from the parabolic form, and different corrections 
have been introduced \cite{GG,PG}. Also, the important role of the combined
effect of the Coulomb forces and spin interaction leading to a
redistribution of single-electron levels was demonstrated
for the whole energy spectrum, especially for the
low-lying single-electron states in the homogeneous magnetic field
\cite{Wag,MC92,Oh}. The ground state transitions predicted 
in this way have been observed experimentally \cite{5}. 

It seems therefore natural to assume that the properties 
of the electron states close to the Fermi level are determined by 
an effective mean field. It is true that the external field is the 
dominant part of the mean field, and thus the effective confining potential 
should reflect the main features of it. Yet it must also
contain the effect of the interplay between Coulomb forces and the external
fields which are governed by the charges in the adjacent layers and
gates and the magnetic field. Due to these considerations, we assume that
the confining potential should also take into
account the changes that affect the properties of the single-electron
states owing to a variation of the homogeneous magnetic field as well as the
slab thickness.

Based on the results discussed in \cite{HK} we conclude that
the harmonic oscillator potential can serve as one of the
phenomenological effective confining potentials in real samples, at 
least for small electron numbers. 
In this paper we study a simple model of a three-dimensional 
quantum dot in an arbitrarily oriented magnetic field for electron numbers 
for which the assumption of a parabolic type potential is still physically 
reasonable. Our procedure invokes an effective dependence
of the parameters of the confining potential on the magnetic field.
In this way, we accommmodate the different mechanisms mentioned 
above and we allow the system to adjust and change its shape under 
variation of the applied magnetic field and the particle number. We assume 
that the electron system behaves like an isolated yet confined droplet 
(see also discussion in \cite{Kast}).
One of the major result is a dependence of the single-electron states on
the slab thickness. Such dependence
shows in the FIR frequencies on which our analysis is focussed. 
We also report on transition strengths in absorption
experiments and on the spatial extension of the confining
potential in a quantum dot. Preliminary results have been reported in
\cite{HN}.

\section{The model}

We consider the Hamiltonian for $N$ noninteracting electrons
moving in an effective mean field, i.e. a three-dimensional
harmonic oscillator
\begin{equation}
H={1\over 2m}\sum_{j=1}^N (\vec  p_j -{e\over c}\vec  A_j)^2+{m\over 2}
\sum_{j=1}^N (\omega _x^2x_{j}^2+\omega _y^2y_{j}^2+\omega _z^2z_{j}^2)
\label{ham}
\end{equation}
with $\vec  A=[\vec  B\times \vec  r]/2$ being the vector potential associated
with the homogeneous magnetic field $\vec  B$. The model does not explicitly 
take into account spin degrees of freedom. In other words, the 
system is assumed to be in a fixed spin state of an arbitrary degree of 
polarisation. Only the orbital motion is affected by the magnetic field.
Using quantum mechanical equations of motion Yip \cite{Yip}
calculated resonance frequencies for this model. Our approach is similar in 
spirit to \cite{Yip}, but we first solve the classical problem, 
from which all quantum mechanical results easily follow. 
In contrast to \cite{Yip,Li} where the FIR frequencies are completely
determined by the external potential, the effective mean field 
responds in our approach to the change of the magnetic field and particle 
number and thus affects the properties of single-particle excitations.
We determine the frequencies
$\omega _i,\,i=x,y,z$ by minimizing the total energy for a given number of
electrons $N$. In this way we allow the system to adjust
the shape of the confining potential under the influence of the applied
magnetic field and the particle number. 
 
The classical equations of motion
for the cartesian components of the momentum and position coordinates read 
\begin{equation}
{d\over dt}\pmatrix{{\vec p} \cr {\vec r}}= {\cal M}
\pmatrix{{\vec p}\cr {\vec r}}  
\label{eqm1}
\end{equation}
where $\vec p$ and $\vec r$ are combined to the six dimensional column vector 
denoted below as $\{ \vec p,\vec r\}$. The matrix 
${\cal M}$ is given by
\begin{eqnarray} &{\cal M}=   \nonumber \\   \nonumber  \\
&\pmatrix{
0&-\Omega _z &\Omega _y &-\omega^2_x-\Omega _y^2-\Omega _z^2
&\Omega _x \Omega _y & \Omega _x \Omega _z\cr
\Omega _z &0&-\Omega _x &\Omega _x \Omega _y
&-\omega^2_y-\Omega _x^2-\Omega _z^2
&\Omega _y \Omega _z \cr
-\Omega _y &\Omega _x &0&\Omega _x \Omega _z&\Omega _y \Omega _z
&-\omega^2_z-\Omega _x^2-\Omega _y^2 \cr
1&0&0&0&-\Omega _z &\Omega _y \cr
0&1&0&\Omega _z &0&-\Omega _x\cr
0&0&1& -\Omega _y &\Omega _x &0 }
\label{matr} \end{eqnarray}
where the notation $\vec \Omega=e\vec B  /( 2mc)$
is used. Denoting the eigenvalues of ${\cal M}$ by
$\pm iE _k,\,k=1,2,3$ we obtain from the form
${\cal M}={\cal U} {\cal D}{\cal V}$ 
the classical solution
\begin{equation}
\{ \vec p(t),\vec r(t)\} =
{\cal U}\exp({\cal D}t){\cal V} \{\vec p(0),\vec r(0)\} 
\end{equation}
where ${\cal D}={\rm diag}(-iE_1,-iE_2,-iE_3,iE_3,iE_2,iE_1)$.
While the classical orbits are of little interest in the present context,
the eigenmodes $E _k$ and the system of eigenvectors listed in the
matrix $\cal U$ are essential for the quantum mechanical treatment.

The eigenmodes which are usually referred to as the normal modes are obtained
from the secular equation 
\begin{equation}
\det|E I-{\cal M}|=0
\end{equation}
with $I$ denoting the six-dimensional unit matrix. The above equation 
turns out to be a third order polynomial in $E ^2$ 
and has also been found by \cite{Yip}. The column
vectors of $\cal U$ are the (complex) right hand eigenvectors of $\cal M$
(note that $\cal M$ is not symmetric, hence neither $\cal U$ nor $\cal V$ are
unitary, yet ${\cal V}= {\cal U}^{-1}$). We denote
the column vectors of $\cal U$ by $u^{(k)}$, they obey the equations
\begin{eqnarray}
({\cal M}+iE _kI)u^{(k)}&=&0,\quad k=1,2,3 \nonumber \\
({\cal M}-iE _{7-k}I)u^{(k)}&=&0,\quad k=4,5,6
\label{eigvec} \end{eqnarray}
which can be solved as an inhomogeneous system by choosing for instance the
sixth component of $u^{(k)}$ equal to unity and then determining the
normalisation as described in Appendix A. With the proper normalisation
the classical Hamilton function of Eq.(\ref{ham}) is cast into the quantum
mechanical form
\begin{equation}
H=\sum _k^3\hbar E _k(Q^{\dagger}_kQ_k + {1\over 2}).
\label{qam}
\end{equation}
The normal mode boson operators $Q^{\dagger}$ and $Q$ are related to
the quantised version of the classical coordinates by
\begin{equation}
\{\vec p,\vec r\}={\cal U} \{Q,Q^{\dagger }\},  \label{lin}
\end{equation}
where we denote by $\{Q,Q^{\dagger }\}$ a column vector which
is the transpose of the vector
$(Q_1,Q_2,Q_3,Q_3^{\dagger },Q_2^{\dagger },Q_1^{\dagger })$.
The properties of the matrix ${\cal U}$ guarantee the commutator
$[Q_k,Q_{k'}^{\dagger }]=i\hbar \,\delta _{k,k'}$ as a consequence of
$[x_i,p_j]=i\hbar \,\delta _{i,j}$. We note that Eq.(\ref{qam}) is the
exact quantised equivalent of Eq.(\ref{ham}).

\section{Quantum energy minimisation}

The total energy of an $N$-particle system associated with the Hamiltonian
Eq.(\ref{qam}) is given by
\begin{equation}
E_{{\rm tot}}=\sum_{j,k} \hbar E _k(n_k+1/2)_j. \label{etot}
\end{equation}
The occupation numbers $n_k$ are the eigenvalues of $Q^{\dagger}_kQ_k$ and
take the values $0,1,2,\ldots $. The ground state is determined by filling
the single-particle levels $\sum_k \hbar E _k(n_k+1/2)$
from the bottom. We take care of the electron spin only 
in obeying the Pauli principle which allows two particles in one level. 
It is clear that different sets of 
normal modes yield different sets of occupation numbers.
The normal modes depend on the three components of the magnetic
field and on the harmonic oscillator frequencies. From our assumption
that the system adjusts itself under the influence of the magnetic
field by minimising $E_{{\rm tot}}$, a variation of the magnetic field
strength leads to a corresponding change of the confining effective potential
which is given by the oscillator frequencies. In other words, for a given
magnetic field, we must seek the minimum of $E_{{\rm tot}}$ under variation
of the oscillator frequencies. The variation cannot be unrestricted as the
confining potential encloses a fixed number of electrons, and assuming that
the electron density does not change we are led to a fixed volume constraint
which translates into the subsidiary condition
$\omega _x\omega _y\omega _z=\omega _0^3$ with $\omega _0$ fixed. Denoting
the Lagrange multiplier by $\lambda $ we solve the variational problem
\begin{equation}
\delta (\langle g|H|g\rangle -\lambda \omega _x\omega _y\omega _z)=0
\label{var} \end{equation}
where $|g\rangle $ denotes the ground state as described above.

From Eq.(\ref{var}) we obtain, after differentiation with respect to the
frequencies and using Feynman's theorem \cite{Fey}
\begin{equation}
{d\over d\omega _k}\langle g|H|g\rangle =
\langle g|{dH\over d\omega _k}|g\rangle ,  \label{feyn}
\end{equation}
the useful condition
\begin{equation}
\omega _x^2\langle g|x^2|g\rangle=\omega _y^2\langle g|y^2|g\rangle
=\omega _z^2\langle g|z^2|g\rangle
\label{cond}
\end{equation}
which must be obeyed at the minimum of $E_{{\rm tot}}$.

In this paper we restrict ourselves to consideration of
a thin slab which extents essentially in two dimensions. This is
achieved by varying only $\omega _x$ and $\omega _y$ in the
minimization procedure while keeping $\omega _z$ fixed at a value which
is, say, five times larger than the other two frequencies. In this case
only $\omega _x^2\langle g| x^2|g\rangle=\omega _y^2\langle g|y^2|g\rangle $
can be fulfilled. Choosing different
(fixed) values of $\omega _z$ allows to study the dependence of the
results on the slab thickness.

It is known from the two-dimensional isotropic harmonic oscillator that
shell closing occurs for Fermions at the 
numbers $2,6,12,20,30,\ldots $. At these numbers strong shell effects
manifest themselves as the most stable quantum configurations.
As a consequence, for $\omega _z\gg \omega _x$
and $B=0$ we must expect the minimum of $E_{{\rm tot}}$ at the symmetric
condition $\omega _x=\omega _y$ for such electron numbers. In the mean 
field approach, the breaking of spherical
symmetry of the potential gives rise to a deformed shape
(Jahn-Teller effect \cite{JT}). For example, rotational spectra of
nuclei and fine structure in the mass spectra of metallic clusters
are explained as a consequence of deformed equilibrium potentials
of these systems (see \cite{BM,Com}).
Therefore, for numbers between the shell numbers a deformed
configuration ($\omega _x\ne \omega _y$) can be expected;
we note that there are always two symmetric solutions: $\omega _x>\omega _y$
and $\omega _x<\omega _y$. The condition
$\omega _z\gg \omega _x$ ensures that we have a genuine two-dimensional
problem in that no particle occupies a quantum mode in the $z$-direction
$(n_z=0)$. However, if the slab is made thicker ($\omega _z$ smaller), the
occupation of the first mode $(n_z=1)$ will occur at the top end of the
occupied levels, which could mean that we find then the symmetric
minimum at $8,14,22,\ldots $, since two particles are occupying the 
$z$-mode that has become available. The
distinction between a slab, that is sufficiently thin so as to
prevent occupation of the first mode in the $z$-direction, and the slab which
can accommodate the first mode is seen physically in the degeneracy of the
two lowest normal modes at $B=0$.
An estimate for $\omega _z^0$ (see Appendix B)
which is the frequency that just
forbids occupation of a $z$-mode is given by $\omega _z^0\ge \omega _{\perp }
(\sqrt{4N+1}-3)/2$ with $\omega _{\perp }$ being the average of $\omega _x$
and $\omega _y$.

\section{Discussion of results}

The level spacing $\hbar \omega _0$ of the oscillator potential 
is determined by equating the Fermi energy $\varepsilon_F$ 
of a free electron gas 
with the potential energy $V(r)={1\over 2}m^{\star }\omega _0 ^2 <r^2>$.
Assuming that the radius of a quantum dot grows with $N^{1/3}$ 
we have  
$<r^2>= 3/5 R_0^2 N^{2/3}$. Consequently,
we have chosen $\hbar \omega _0=1.35N^{-1/3} \epsilon _{{\rm F}}$ where the
Fermi energy is obtained from the mean radius
$R_0$ and effective mass of typical quantum dots; for GaAs, $R_0$=320\AA \
and $m^{\star }=0.067m_e$ yield $\epsilon _{{\rm F}}=2$meV. Throughout the
paper we use meV as energy units, \AA \ for length and Tesla for the
magnetic field strengths.

\subsection{The spectrum}

The three normal modes which can be discerned experimentally
as excitation energies in FIR-spectroscopy \cite{6}
behave in a distinctly characteristic way when the magnetic field is
switched on. We recall that the values for $\omega _x$ and $\omega _y$
are fixed by minimizing $E_{{\rm tot}}$.
In Fig. 1 we display typical patterns for two different
electron numbers. In both cases, $\omega _z$ is chosen so large that
$n_3$ remains zero. Note that, when the magnetic field is switched on, the
occupation numbers $n_1,n_2,n_3$ refer to the normal modes and can no longer
necessarily be associated with an $x$-, $y$- or $z$-direction.

\vspace*{-0.1cm}
\begin{figure}
\epsfxsize=2.8in
\centerline{
\epsffile{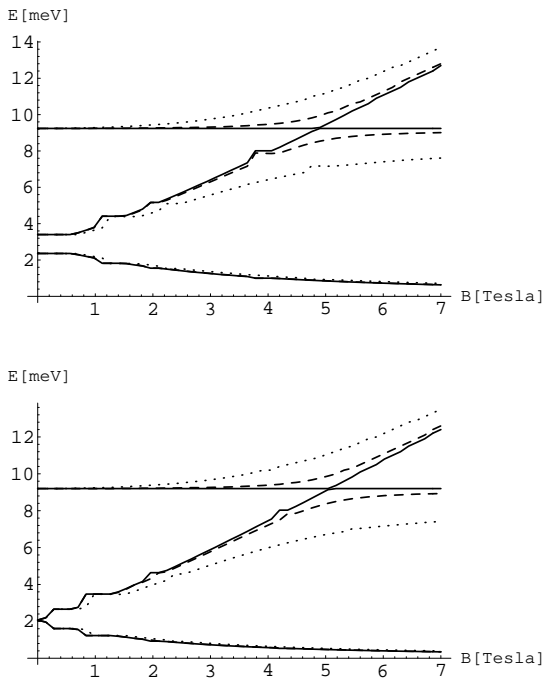}
}
\vglue 0.2cm
\caption{ 
Normal mode frequencies as a function of magnetic field strength
for $N=16$ (top) and $N=30$ (bottom). Solid, dashed and dotted lines
illustrate results for $\theta =0^0,10^0$ and $30^0$, respectively. Values for
$\hbar \omega _z$ are 9.2 in all cases.
}
\label{fig1}
\end{figure}

The degeneracy for $N=30$ is clearly
seen at $B=0$ whereas three distinct modes, corresponding to a deformed
shape, are seen at $B=0$ for $N=16$. The behaviour is shown for values of
$\theta =0^0,10^0$ and $30^0$. For $\theta =0$ the third mode
(the highest at $B=0$) does not
interact with the applied field, therefore there is a level crossing at
about $B=5$. This becomes a level repulsion for $\theta \ne 0$ as now
all modes are affected by the magnetic field. As a consequence, the
$\theta$-dependence is strongly pronounced for the two upper modes, while
such dependence is insignificant for the lowest mode. Here we find
quite naturally an interpretation of avoided level crossing observed in
the experiments \cite{6} associated with shape variations
in a tilted magnetic field. We have also investigated
the $\phi $-dependence of the normal modes, which is expected to become
significant only for large values of $\theta $; the differences are small,
they can not be seen in Fig. 1. Below we return
to a particular situation where the $\phi $-dependence is important.

\vspace*{-0.1cm}
\begin{figure}
\epsfxsize=2.8in
\centerline{
\epsffile{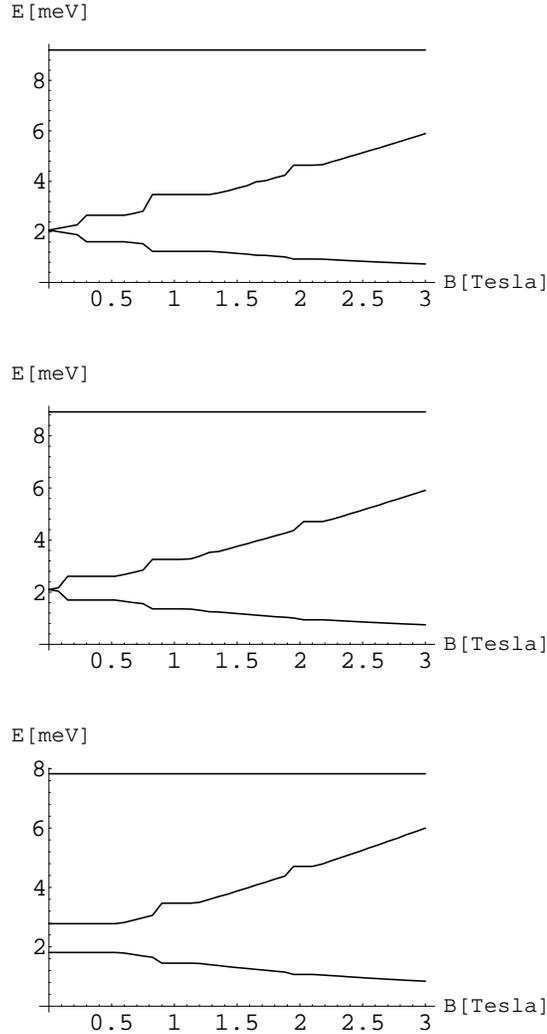}
}
\vglue 0.2cm
\caption{ 
Normal mode frequencies for $N=30$ and $\theta =0$. The slab
thickness increases from the top to the bottom.
}
\label{fig2}
\end{figure}

There are windows of magnetic field strength for which the normal modes are
virtually constant. They are discernible especially at smaller values of the
magnetic field strength and are a reflection of shape changes. For $N=30$
the system is rotationally symmetric at $B=0$. At $B>0.2$ it changes into
a deformed shape ($\omega _x\ne \omega _y$) which is associated
with the larger derivatives of the two
lower normal modes. The symmetric shape ($\omega _x=\omega _y$) is now
being restored in a continuous fashion when $B$ is further increased, the two
normal modes are virtually constant during this smooth transition. The
constancy of the normal modes holds strictly only for $\theta =0$ and can be
shown analytically; this is deferred to Appendix C. At $B=0.7$,
where the system has eventually re-gained its symmetric shape, a second
shape transition occurs for $B>0.7$, which is reflected by the second sharp
decrease/increase of the first/second normal mode. This pattern continues,
yet it is less and less pronounced with increasing field strength. It remains
to be seen whether these shape changes can be dissolved experimentally;
finite temperature could disturb this pattern.

Making the slab thicker, that is choosing $\omega _z$ smaller so as to allow
occupation of the next level of the third mode ($n_3=1$), invokes distinct
changes which could possibly be dissolved experimentally. In Fig. 2 we
display the effect of varying the slab thickness for $N=30$. For clarity we
focus on smaller magnetic field strength and we display only results for
$\theta =0$. The top figure is to facilitate comparison; the results are a
repeat of Fig.(1b). In the middle figure the slab thickness has been
decreased in such a way that the third mode becomes occupied when the field
is switched on while it is unoccupied for zero field strength. As a
consequence, the shape transition occurs for smaller field strength, it is
here directly associated with the switch from $n_3=0$ to $n_3=1$ which is
induced by the magnetic field in this particular case. The bottom figure
refers to the case where $n_3=1$ throughout. The major effect is the lifting
of degeneracy for $\vec B=0$. Since there are effectively only 28 particles
occupying the two lower modes, the confining potential is deformed.

With increasing field strength the potential changes into a symmetric
shape which is attained at about
$|\vec B|=0.55$ where it again undergoes a transition to a deformed shape in
line with the discussion above. In this context we note that an increase of
$\omega _z$ beyond the values used in the results presented, which means an
even thinner slab, pushes the upper level further away. A truly
two-dimensional setting should therefore yield only two observable levels.

\vspace*{-0.1cm}
\begin{figure}
\epsfxsize=2.8in
\centerline{
\epsffile{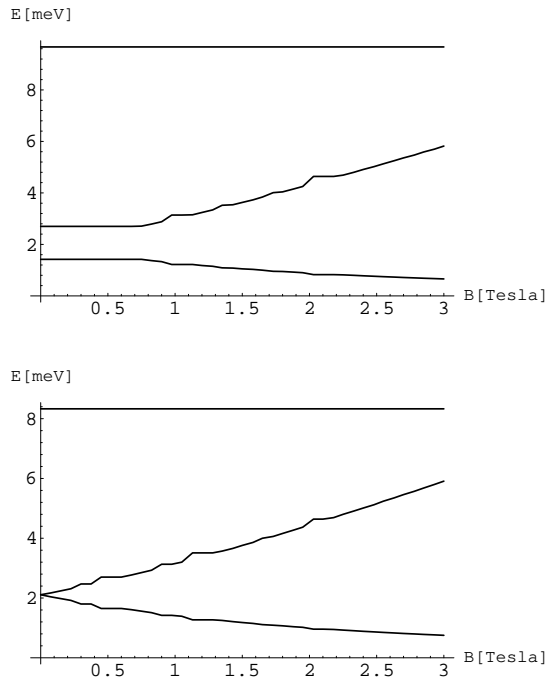}
}
\vglue 0.2cm
\caption{ 
Normal mode frequencies for $N=32$ and $\theta =0$. The slab
thickness increases from the top to the bottom.
}
\label{fig3}
\end{figure}

The effective occupation in the two lower modes can also be demonstrated
by using $N=22$ or $N=32$. For $n_3=1$ this yields the effective numbers
20 or 30 in the planary motion with the expectation that the energy minimum
is adopted for a symmetric shape. This is nicely confirmed and demonstrated
in Fig. 3 where results for $N=32$ are presented. The top and bottom figures
illustrate the cases where $n_3=0$ and $n_3=1$, respectively. The
interpretation is selfexplanatory after the previous discussions.

In all the examples we see quantum mechanics nicely at work. Despite the
simplicity of the model, these effects should be seen in an experiment.

\subsection{Dipole transitions}

The actual observation of the three modes is achieved by electromagnetic
transitions, for instance in absorption experiments.
Within the single-particle model considered here transition matrix elements
between the ground state and single-particle excitations are available for 
any single-particle operator. The electric dipole transition is expected to
feature prominently. We note that this transition rules out a spin flip,
which is in line with our assumption about fixed spin states.
We interpret the three observed frequencies \cite{6}
as the three normal modes. If no photon polarisation is measured in the
initial and final states, the dipole transition strength is given by
\begin{eqnarray}
S&=& \sum _{f,m_f,m_{\gamma }}|\langle f|E1|g\rangle |^2 \nonumber \\
&=& \sum _{f,m_f} (
|\langle f|x|g\rangle|^2+|\langle f|y|g\rangle|^2+|\langle f|z|g\rangle|^2),
\label{dip} \end{eqnarray}
since the electric dipole operator is proprtional to $\vec r$.
Using Eq.(\ref{lin}) the matrix elements are given by a sum over the
appropriate entries of the matrix $\cal U$. The selection rules allow only
transitions with $n_k\to n_k\pm 1$ by which all accessible final states are
determined. Note that the configuration $\{n_k\}$ is different for different
magnetic field strength. We list some results in Table 1. They refer to
$\theta =0$ and $n_3=0$ for the ground state. These figures do change as a
function of $\theta $, electron number and slab thickness. The results
demonstrate that there is a redistribution of dipole strength due to
variation of these parameters and therefore of the shape of the system.
This aspect could be the subject of further experimental analysis.

\begin{center}
\begin{tabular} {|c||l r|l r|l r|} \hline
 & $S$ & $E_1 $ & $S$ & $E_2 $ & $S$ & $E_3 $ \\ \hline \hline
$B=0.1$ & 1.0& 2.0 &1.0& 2.1 & 0.7 &9.2 \\  \hline
$B=1$ & 0.7& 1.2 &1.2& 3.5 & 0.7 &9.2  \\  \hline
$B=7$ & 0.07& 0.3 &0.7& 9.2 & 1.1& 12.4  \\ \hline
\end{tabular}
\end{center}
{\bf Table 1.} Dipole strengths $S$ for various values of the magnetic field.
The associated frequencies are indicated next to the strengths.
The numbers are normalised to the lowest frequency transitions at zero field.

\subsection{Mean square values}

The calculation of matrix elements of $x^2$ and $y^2$ which are
characterized by the shape of the effective confining potential 
follows similar lines
as for the dipole matrix elements. In Table 2 we give a few values for
$N=30$ and $\theta =0$. Note the strong deformation at $B=0.8$.
For larger field strengths the mean values decrease, since the effective
frequencies of the confining potential are augmented by the Larmor frequency;
this effect also enhances a symmetric shape to an increasing extent for
increasing field strength. We stress that these values all obey the
condition Eq.(\ref{cond}). For
$\theta >0$ the pattern remains essentially the same except
for the fact, that the shape is not symmetric for large field strength but
deformed; the degree of deformation depends on $\theta $.
This is again understood from the Larmor frequencies which are now different
for the different directions. This argument is nicely confirmed when
using $\theta >0$ and $\phi =45^0$; now the symmetry is guaranteed by
the choice $B_x=B_y$, and, in fact, the system settles down to a symmetric
configuration for large field strength.
For $B=7,\theta =30^0,\phi=0$ the mean values are 1015 and 1069, respectively,
while for $\phi =45^0$ they are both equal to 1044. Note that such pattern
for large field strength is independent of the particle number. For instance,
irrespective of the strongly deformed shape for $N=16$ at $B=0$,
we find at $B=7$ also a symmetric shape for $\phi =45^0$ but
a slight deformation for $\theta =30^0$ and $\phi =0$. Again we stress that
the figures depend on the slab thickness in the spirit of the previous
paragraph.
\begin{center}
\begin{tabular} {|c||c|c|} \hline
 & $\sqrt{\langle g|x^2|g\rangle }$ & $\sqrt{\langle g|y^2|g\rangle }$ \\
 \hline   \hline
 $B=0$ & 1167 & 1167 \\ \hline
 $B=0.8$ & 770 & 1773 \\ \hline
 $B=7$ & 981 & 981 \\ \hline
\end{tabular}
\end{center}
{\bf Table 2.} Mean square values of the extension in the $x-y-$plane of the
quantum dot under variation of the magnetic field strength.

\section{Summary}

In view of the good qualitative
agreement with experimental data \cite{6} we think that the model gives
a fair account about the essential features of quantum dots on semiconductor
interfaces. The simplicity of the soluble model allows consideration of an
arbitrarily oriented magnetic field. The minimal energy requirement invokes
a re-arrangement of the quantum configuration which is manifested as
shell effects and hence shape changes of the quantum dot. By this mechanism 
we find an effective
$N$-dependence of the FIR frequencies in quantum dots (see Fig.$\,$1).
An interesting aspect of the model is the constancy of the FIR frequencies
(the normal modes) under variation of the magnetic field as long as the
potential is deformed. As is discussed in Section 4 the system responds to
the increasing magnetic field strength by changing the confining potential
towards a spherical shape rather than by changing its energy. This is shown
analytically in Appendix C for $\theta =0$ where it strictly holds. It is
interesting to note that for $\theta \ne 0$ the statement still holds
to high accuracy. As long as the shape of the effective
confining potential is similar for different electron numbers, the
splitting of the FIR frequencies due to the magnetic field should be equally
alike. 
When the tilting angle $\theta$ is switched on, all three modes are globally
affected by the magnetic field leading to an avoided level
crossing or anticrossing \cite{6}. A thorough experimental analysis of a
variation of slab thickness, tilted angle and strength of the magnetic field,
which leads to the manifestation of quantum effects such as shape variations 
for fixed and different numbers of electrons, could assess the reality
and limitations of the model.

\vspace{0.4in}
The authors acknowledge financial support from the Foundation for Research
Development of South Africa which was provided under the auspices of the
Russian/South African Agreement on Science and Technology. R.G.N. is thankful
for the warm hospitality which he experienced from the whole Department of
Physics during his visit.

\vspace{0.4in}
{\it Note added in proof:}\par

After submission of the present paper a report about recent experiments,
 which confirm essential aspects of this investigation,
has appeared in Physical Review Letters, Vol. 77, 3613 (1996) by
S. Tarucha, D.G. Austing, T.Honda, R.J. van der Hage and L.P. Kouwenhoven.

\appendix
\def\theequation{\thesection.\arabic{equation}}
\section{Normalisation of the eigenvectors}
\setcounter{equation}{0}
The Hamilton function of Eq.(\ref{ham}) can be written in matrix form
\begin{equation}
H=\{ \vec p,\vec r \}^T{\cal H} \{ \vec p,\vec r \}
\end{equation}
where
\begin{equation}
{\cal H}=\pmatrix{1&0&0&0&-\Omega _z&\Omega _y \cr
0&1&0&\Omega _z&0&-\Omega _x \cr
0&0&1&-\Omega _y&\Omega _x&0 \cr
0&\Omega _z&-\Omega _y&
\omega _x^2+\Omega _y^2+\Omega _z^2&-\Omega _x\Omega _y&-\Omega _x\Omega _z\cr
-\Omega _z&0&\Omega _x&-\Omega _x\Omega _y&
\omega _y^2+\Omega _z^2+\Omega _x^2&-\Omega _y\Omega _z\cr
\Omega _y&-\Omega _x&0&-\Omega _x\Omega _z&-\Omega _y\Omega _z &
\omega _z^2+\Omega _x^2+\Omega _y^2 }
 \end{equation}
We aim at the quantum mechanical form
\begin{equation}
H= \{Q,Q^{\dagger }\}^T {\cal H}_{qm} \{Q,Q^{\dagger }\}    
\label{bos}
\end{equation}
where
\begin{equation}
{\cal H}_{qm}=\pmatrix{0&0&0&0&0&E _1 \cr
0&0&0&0&E _2&0 \cr
0&0&0&E _3&0&0 \cr
0&0&E _3&0&0&0 \cr
0&E _2&0&0&0&0 \cr
E _1&0&0&0&0&0 }.   \label{skew}
\end{equation}
Exploiting the fact that
\begin{equation}
{\cal M}=\pmatrix {0&-I\cr I&0}{\cal H}
\end{equation}
where $I$ is a 3 by 3 unit matrix, it follows that, up to normalisation
factors, the matrix ${\cal V}={\cal U}^{-1}$ can be
written as
\begin{equation}
{\cal V}=\pmatrix{0&0&0&0&0&-i \cr
0&0&0&0&-i&0 \cr
0&0&0&-i&0&0 \cr
0&0&i&0&0&0 \cr
0&i&0&0&0&0 \cr
i&0&0&0&0&0 }{\cal U}^T\pmatrix {0&-I\cr I&0}.
\end{equation}
This implies that ${\cal U}^T \cal H U$ is in fact skew-diagonal as
in Eq.(\ref{skew}), and therefore $\cal U$ can be normalised such that
${\cal U}^T {\cal H U}={\cal H}_{qm}$.  Using 
Eq.(\ref{lin}), Eq.(\ref{bos}) follows.

\section{Estimate of $\omega_z^0$}
\setcounter{equation}{0}

Consider the energy of the three-dimensional oscillator
\begin{equation}
E = \hbar \omega_z (n_z + 1/2) + \hbar \omega_\perp (n_\perp + 1) =
E_z + E_\perp
\end{equation}
where $\omega_x = \omega_y = \omega_\perp \neq \omega_z$.
The condition $\omega _z\gg \omega _\perp$ ensures that we have a
genuine two dimensional
problem in that no particle occupies a quantum mode in the $z$-direction
$(n_x \neq 0, n_y \neq 0, n_z = 0)$, i.e. the first level with
$(n_x = 0, n_y = 0, n_z = 1)$ lies higher than the
ones which are filled for given particle number. From the condition
$E^0(n_x=n_y=0, n_z=1)\geq E(n_x + n_y = n_\perp \neq 0, n_z = 0)$
it follows
\begin{equation}
\label{wz}
\omega_z^0 \geq n_\perp\omega_\perp
\end{equation}
For a two-dimensional system the number of particles
(two particles per level)
is $A = (N+1)(N+2)$ where $N = n_x + n_y = n_\perp$ is
the shell number of the last filled shell. Therefore, from the equations
above it follows
\begin{equation}
\label{nperp}
A=(n_\perp +1)(n_\perp +2) \quad {\rm i.e.}\quad
 n_\perp = \frac{\sqrt{4A+1}-3}2
 \end{equation}
 Substituting Eq.(\ref{nperp}) into Eq.(\ref{wz}), we obtain
 \begin{equation}
 \omega_z^0 \geq {{\omega_\perp}\over 2} (\sqrt{4A+1}-3)
 \end{equation}

\section{Constancy of the normal modes}
\setcounter{equation}{0}

For $\theta =0$ the normal modes are
\begin{eqnarray}
E _1^2&=&{1\over 2}(\omega _x^2+\omega _y^2+4\Omega _z^2+
\sqrt{(\omega _x^2-\omega _y^2)^2+8\Omega _z^2(\omega _x^2+\omega _y^2)
 + 16\Omega _z^4})\nonumber\\
\label{sol}
E _2^2&=&{1\over 2}(\omega _x^2+\omega _y^2+ 4\Omega _z^2-
\sqrt{(\omega _x^2-\omega _y^2)^2+8\Omega _z^2(\omega _x^2+\omega _y^2)
+ 16\Omega _z^4})\\
E _3&=&\omega _z.\nonumber
\end{eqnarray}
The last equation shows that the largest mode $E _3$ is independent
of the magnetic field strength, since $\omega _z$ is kept fixed. From the
first two equations we obtain a relation between two sets of frequencies
$\omega_{1x},\omega_{1y}$ and $\omega_{2x},\omega_{2y}$ which yield the same
normal modes for different field strength. With the notation
$\Delta ^2=\Omega_{2z}^2-\Omega_{1z}^2>0$, where $\Omega_{1z}$
and $\Omega_{2z}$
refer to the different field strengths $B_1$ and $B_2$, we obtain
from the requirements $E _{1,2}(B_1)=E _{1,2}(B_2)$
\begin{eqnarray}
\label{omeg}
\omega_{2x}^2&=&{1\over 2}(\omega_{1x}^2+\omega_{1y}^2-4\Delta ^2+
\sqrt{(\omega_{1x}^2-\omega_{1y}^2)^2-8\Delta ^2(\omega_{1x}^2+\omega_{1y}^2)
 + 16\Delta ^4})\nonumber\\
    \\
\omega_{2y}^2&=&{1\over 2}(\omega_{1x}^2+\omega_{1y}^2- 4\Delta ^2-
\sqrt{(\omega_{1x}^2-\omega_{1y}^2)^2-8\Delta ^2(\omega_{1x}^2+\omega_{1y}^2)
+ 16\Delta ^4})     \nonumber
\end{eqnarray}
To ensure that $\omega_{2x}$ and $\omega_{2y}$ are real, $\Delta $ must
obey the condition
\begin{equation}
\sqrt{\Delta } \le {|\omega _{1x}-\omega _{1y}|\over 2}.   \label{ineq}
\end{equation}
This means that
the normal modes {\it must} change with the magnetic field strength for a
spherical shape ($\omega_{1x}=\omega_{1y}$) as is established by the results.
In turn, it is possible that $E _1$ and $E _2$ do not change under
variation of $B$ as long as $\omega_{1x}\ne \omega_{1y}$. Further, if the
magnetic field strength has increased up to the value where
condition (\ref{ineq}) becomes an equality, then the spherical shape
($\omega_{2x}=\omega_{2y}$) is
attained. Recall that throughout this transition from deformed to spherical
shape the normal modes (Eq.(\ref{sol})) have not changed.

It remains to show that this solution is in fact the minimal energy
solution. For $\theta =0$ the explicit expressions for
$\langle g|x^2|g\rangle $ and $\langle g|y^2|g\rangle $ read
\begin{eqnarray}
\label{til}
\langle g|x^2|g\rangle  = {1\over {2m}}\left[\frac{\Sigma_1}{E_1} +
\frac{\Sigma_2}{E_2}
+ \frac{\omega_x^2 -\omega_y^2 + 4\Omega_z^2}{E_1^2 - E_2^2}
\left(\frac{\Sigma_1}{E_1} -
\frac{\Sigma_2}{E_2}\right)\right]\nonumber\\
     \\
\langle g|y^2|g\rangle  = {1\over {2m}}\left[\frac{\Sigma_1}{E_1} +
\frac{\Sigma_2}{E_2}
+ \frac{\omega_y^2 -\omega_x^2 + 4\Omega_z^2}{E_1^2 - E_2^2}
\left(\frac{\Sigma_1}{E_1} -
\frac{\Sigma_2}{E_2}\right)\right]\nonumber
\end{eqnarray}
where $\Sigma_k =\sum_j (n_k + {1\over 2})_j$ denotes the sum
over all occupied single-particle levels.
Suppose the condition Eq.(\ref{cond}) is fulfilled for the confining 
frequencies $\omega_{1x}$ and $\omega_{1y}$ for the value of the magnetic 
field $B_1 (\Omega_{1z})$ . Substituting Eqs.(\ref{til})
into Eq.(\ref{cond}) and taking into account that
$\omega_{1x}\neq\omega_{1y}$, we obtain
\begin{equation}
\label{qur}
\left[\frac{\Sigma_1}{E_1} +
\frac{\Sigma_2}{E_2}
+ \frac{\omega_{1x}^2 +\omega_{1y}^2 + 4\Omega_{1z}^2}
{E_1^2 - E_2^2}
\left(\frac{\Sigma_1}{E_1} -
\frac{\Sigma_2}{E_2}\right)\right] = 0.
\end{equation}
From Eqs.(\ref{omeg}) it follows
\begin{equation}
\label{pol}
\omega_{2x}^2 +\omega_{2y}^2 = \omega_{1x}^2 +\omega_{1y}^2 -
4(\Omega_{2z}^2-\Omega_{1z}^2).
\end{equation}
Using the result of Eqs.(\ref{til}),(\ref{qur}) and (\ref{pol}),
and the fact that $\Sigma_1$  and $\Sigma_2$ remain the same, we obtain
\begin{eqnarray}
\omega_{2x}^2\langle g2|x^2|g2\rangle  -
\omega_{2y}^2\langle g2|y^2|g2\rangle = \nonumber\\
(\omega_{2x}^2 - \omega_{2y}^2)
\left[\frac{\Sigma_1}{E_1} +
\frac{\Sigma_2}{E_2}
+ \frac{\omega_{2x}^2 +\omega_{2y}^2 + 4\Omega_{2z}^2}
{E_1^2 - E_2^2}
\left(\frac{\Sigma_1}{E_1} -
\frac{\Sigma_2}{E_2}\right)\right] =\nonumber\\
(\omega_{2x}^2 - \omega_{2y}^2)
\left[\frac{\Sigma_1}{E_1} +
\frac{\Sigma_2}{E_2}
+ \frac{\omega_{1x}^2 +\omega_{1y}^2 + 4\Omega_{1z}^2}
{E_1^2 - E_2^2}
\left(\frac{\Sigma_1}{E_1} -
\frac{\Sigma_2}{E_2}\right)\right]\equiv 0
\end{eqnarray}
where $|g2\rangle $ denotes the ground state referring to the frequencies 
$\omega_{2x},\omega_{2y}$ and the magnetic field $B_2$. This means that
Eq.(\ref{cond}) is fulfilled.


\newpage

\begin{thebibliography}{99}
\bibitem{TC} T.Chakraborty, Comments Condens. Matter Phys. {\bf 16} 35 (1992).
\bibitem{MK} M.A. Kastner, Rev.Mod.Phys. {\bf 64}, 849 (1992).
\bibitem{Jon} N.F. Johnson, J.Phys.: Condens.Matter {\bf 7} 965 (1995).
\bibitem{BM} A. Bohr and B. R. Mottelson, {\it Nuclear Structure}
(Benjamin, New York, 1975), Vol.2.
\bibitem{Ni} H. Nishioka, K. Hansen, and B. R. Mottelson, Phys. Rev.
{\bf B42}, 9377 (1990).
\bibitem{He93} W. A. de Heer, Rev. Mod. Phys. {\bf 65}, 611 (1993);
M. Brack, {\it ibid} {\bf 65}, 677 (1993).
\bibitem{HN95}  W.D. Heiss and R.G. Nazmitdinov, Phys. Rev. Lett.{\bf 73},
1235 (1994); W.D. Heiss, R.G. Nazmitdinov and S. Radu, Phys. Rev.
{\bf B51}, 1874 (1995); W.D. Heiss, R.G. Nazmitdinov and S. Radu, Phys. Rev.
{\bf C52}, 3032 (1995).
\bibitem{Com} Comments At.Mol.Phys. {\bf 31} Nos. 3-6 (1995).
\bibitem{1} Ch.Sikorski and U.Merkt,  Phys.Rev.Let. {\bf 62}, 2164 (1989).
\bibitem{2} W.Hansen, T.P. Smith III, K.Y. Lee, J.A. Brum, C.M. Knoedler,
J.M. Hong and D.P. Kern, Phys.Rev.Lett. {\bf 62}, 2168 (1989).
\bibitem{3} T. Demel, T. Heitmann, P. Grambow and K. Ploog,
Phys.Rev.Lett. {\bf 64}, 788 (1990).
\bibitem{4} P.L. McEuen, E.B. Foxman, U. Meirav, M.A. Kastner, Yigal Meir,
Ned S. Wingreen and S.J.Wind, Phys.Rev.Lett. {\bf 66}, 1926 (1991).
\bibitem{5} R.C. Ashoori, H.L. Stormer, J.S. Weiner, L.N. Pfeiffer,
K.W.Baldwin and K.W. West, Phys.Rev.Lett. {\bf 71}, 613 (1993).
\bibitem{6} B.Meurer, D.Heitmann and K.Ploog, Phys.Rev. {\bf B48}, 11488
(1993).
\bibitem{BJH} L.Brey, N.F. Johnson and B.I. Halperin, Phys.Rev. {\bf B40},
10647 (1989).
\bibitem{P} F.M. Peeters, Phys.Rev. {\bf B42}, 1486 (1990).
\bibitem{MC} P.A. Maksym and T. Chakraborty, Phys.Rev.Lett. {\bf 65}, 108
(1990).
\bibitem{Fock} V.Fock, Z.Phys. {\bf 47} 446 (1928);
C.G. Darwin, Proc.Cambridge Philos. Soc. {\bf 27}, 86 (1930).
\bibitem{Kohn} W. Kohn, Phys. Rev. {\bf 123}, 1242 (1961).
\bibitem{Bak} P.Bakshi, D.A. Broido and K. Kempa, Phys.Rev. {\bf B42}, 7416
(1990).
\bibitem{Yip} S.K. Yip, Phys.Rev.{\bf B43}, 1707 (1991).    
\bibitem{Li} Q.P. Lie, K.Karrai, S.K. Yip, S.Das Sarma and H.D. Drew,
Phys.Rev. {\bf B43}, 5151 (1991).
\bibitem{KLS} A. Kumar, S.E. Laux and F. Stern, Phys.Rev. {\bf B42},
 5166 (1990).
\bibitem{GG} V. Gudmudsson and R.R. Gerhardts, Phys.Rev. {\bf B43}, 12098 
(1991).
\bibitem{PG} D. Pfannkuche and R.R. Gerhardts, Phys.Rev. {\bf B44}, 13132
(1991).
\bibitem{Wag} M. Wagner, U. Merkt and A.V. Chaplik, Phys.Rev. {\bf B45},
1951 (1992).
\bibitem{MC92} P.A. Maksym and T.Chakraborty, Phys.Rev. {\bf B45}, 1947 (1992).
\bibitem{Oh} J.H.Oh, K.J.Chang, G.Ihm and S.J.Lee, Phys.Rev. {\bf B50},
15397 (1994).
\bibitem{HK} D. Heitmann and J. Kotthaus, Phys.Today {\bf 46}, 56 (1993).
\bibitem{Kast} M.A. Kastner, Comments Condens. Matter Phys.
{\bf 16}, 349 (1996).
\bibitem{HN} W.D. Heiss and R.G. Nazmitdinov, Phys.Lett. {\bf A 222}, 309 
(1996).
\bibitem{Fey} R.P. Feynman, Phys.Rev. {\bf 56}, 340 (1939).
\bibitem{JT} H.A. Jahn and E. Teller, Proc. Roy. Soc. {\bf A161}, 220 (1937).
\end{thebibliography}
\end{document}